\documentclass[
  journal=pasa,
  manuscript=research-paper,
  year=????,
  volume=??,
]{cup-journal}

\usepackage{microtype,siunitx,booktabs}
\usepackage{natbib}
\setcitestyle{aysep={}}

\usepackage{CJKutf8}    
\usepackage{tikz,xcolor,hyperref} 
\usepackage{amsmath}    

\definecolor{lime}{HTML}{A6CE39}
\DeclareRobustCommand{\orcidicon}{%
    \begin{tikzpicture}
    \draw[lime, fill=lime] (0,0) 
    circle [radius=0.16] 
    node[white] {{\fontfamily{qag}\selectfont \tiny ID}};
    \draw[white, fill=white] (-0.0625,0.095) 
    circle [radius=0.007];
    \end{tikzpicture}
    \hspace{-1mm}
}

\newcommand{\orcidChrisO}{\href{https://orcid.org/0000-0003-0017-349X}{\orcidicon}}
\newcommand{\orcidChrisW}{\href{https://orcid.org/0000-0002-4569-016X}{\orcidicon}}
\newcommand{\orcidSamuel}{\href{https://orcid.org/0000-0001-9372-4611}{\orcidicon}}
\newcommand{\orcidPatrick}{\href{https://orcid.org/0000-0003-4237-0520}{\orcidicon}}
\newcommand{\orcidAdrian}{\href{https://orcid.org/0000-0003-4827-9402}{\orcidicon}}
\newcommand{\orcidJack}{\href{https://orcid.org/0000-0001-8359-2328}{\orcidicon}}
\newcommand{\orcidJeno}{\href{https://orcid.org/0000-0003-2835-0304}{\orcidicon}}
\newcommand{\orcidJuan}{\href{https://orcid.org/0000-0002-2647-4373}{\orcidicon}}
\newcommand{\orcidRajeev}{\href{https://orcid.org/0000-0001-7633-7038}{\orcidicon}}
\newcommand{\orcidXiaohui}{\href{https://orcid.org/0000-0003-3310-0131}{\orcidicon}}
\newcommand{\orcidFuyan}{\href{https://orcid.org/0000-0002-1620-0897}{\orcidicon}}

\newcommand\ion[2]{\text{#1\,\textsc{\lowercase{#2}}}}	
\newcommand\arcsec{$^{\prime\prime}$}

\title{Discovery of the most luminous quasar of the last 9~Gyr}

\author{Christopher A.\ Onken}
\affiliation{\orcidChrisO
Research School of Astronomy and Astrophysics, Australian National University, Canberra ACT 2611, Australia}
\email[Christopher A.\ Onken]{christopher.onken@anu.edu.au}

\author{Samuel Lai\begin{CJK}{UTF8}{gbsn}(赖民希)\end{CJK}}
\affiliation{\orcidSamuel 
Research School of Astronomy and Astrophysics, Australian National University, Canberra ACT 2611, Australia}

\author{Christian Wolf}
\affiliation{\orcidChrisW
Research School of Astronomy and Astrophysics, Australian National University, Canberra ACT 2611, Australia;\\
\protect \hphantom{~~~~~~~~} Centre for Gravitational Astrophysics, Australian National University, Canberra ACT 2600, Australia}

\author{Adrian B.\ Lucy}
\affiliation{\orcidAdrian 
Space Telescope Science Institute, 3700 San Martin Drive, Baltimore, MD 21218, USA}

\author{Wei Jeat Hon}
\affiliation{\orcidJack 
School of Physics, University of Melbourne, Parkville, Victoria 3010, Australia}

\author{Patrick Tisserand}
\affiliation{\orcidPatrick 
Sorbonne Universit\'{e}s, UPMC Univ Paris 6 et CNRS, Institut d'Astrophysique de Paris, 98 bis bd Arago, F-75014 Paris, France}

\author{Jennifer L.\ Sokoloski}
\affiliation{\orcidJeno
Columbia Astrophysics Lab 550 W120th Street, 1027 Pupin Hall, MC 5247, Columbia University, New York, NY 10027, USA}

\author{Gerardo J.\ M.\ Luna}
\affiliation{\orcidJuan
CONICET-Universidad de Buenos Aires, Instituto de Astronom\'{i}a y F\'{i}sica del Espacio (IAFE), Av. Inte. G\"{u}iraldes 2620, C1428ZAA, Buenos Aires, Argentina;\\ \protect \hphantom{~~~~~~~~} Universidad de Buenos Aires, Facultad de Ciencias Exactas y Naturales, Buenos Aires, Argentina;\\
\protect \hphantom{~~~~~~~~} Universidad Nacional de Hurlingham, Av. Gdor. Vergara 2222, Villa Tesei, Buenos Aires, Argentina}

\author{Rajeev Manick}
\affiliation{\orcidRajeev
Univ. Grenoble Alpes, CNRS, IPAG, 38000 Grenoble, France;\\
\protect \hphantom{~~~~~~~~} South African Astronomical Observatory, PO Box 9, Observatory 7935, South Africa}

\author{Xiaohui Fan}
\affiliation{\orcidXiaohui
Steward Observatory, University of Arizona, 933 North Cherry Avenue, Tucson, AZ 85721, USA
}

\author{Fuyan Bian \begin{CJK}{UTF8}{gbsn}(边福彦)\end{CJK}}
\affiliation{\orcidFuyan
European Southern Observatory, Alonso de C\'{o}rdova 3107, Casilla 19001, Vitacura, Santiago 19, Chile
}

\doi{}

\received {dd Mmm YYYY}
\revised  {dd Mmm YYYY}
\accepted {dd Mmm YYYY}
\published{dd Mmm YYYY}


\begin{document}

\begin{abstract}
We report the discovery of a bright ($g = 14.5$~mag (AB), $K = 11.9$~mag (Vega)) quasar at redshift $z=0.83$ -- the optically brightest (unbeamed) quasar at $z>0.4$. SMSS~J114447.77-430859.3, at a Galactic latitude of $b=+18.1^{\circ}$, was identified by its optical colours from the SkyMapper Southern Survey (SMSS) during a search for symbiotic binary stars. Optical and near-infrared spectroscopy reveals broad \ion{Mg}{ii}, H$\beta$, H$\alpha$, and Pa$\beta$ emission lines, from which we measure a black hole mass of $\log_{10} (M_{\rm BH}/$M$_{\odot}) = 9.4 \pm 0.5$. With its high luminosity, $L_{\rm bol} = (4.7\pm1.0)\times10^{47}$~erg~s$^{-1}$ or $M_{i}(z=2) = -29.74$~mag (AB), we estimate an Eddington ratio of $\approx 1.4$. As the most luminous quasar known over the last $\sim$9~Gyr of cosmic history, having a luminosity $8\times$ greater than 3C~273, the source offers a range of potential follow-up opportunities.
\end{abstract}

\section{INTRODUCTION }

The observational study of quasars took off rapidly from the back-to-back-to-back papers of {\it Nature}'s 16 March 1963 issue, which featured the redshift determinations for 3C~273 and 3C~48 \citep{1963Natur.197.1040S,1963Natur.197.1040O,1963Natur.197.1041G}. Barely two years later, the quasar 3C~9 became the first known object at a redshift greater than 2 \citep{1965ApJ...141.1295S}. But as exemplified by the 5-magnitude difference in the optical brightness between 3C~273 and 3C~9, the exploration towards higher redshifts became a push to fainter magnitudes.

Fortunately, the early recognition of radio-quiet quasars becoming prominent amongst blue, star-like objects beyond a magnitude of $V=14.5$ (Vega) provided an efficient means of identifying quasars from photometric techniques \citep{1965ApJ...141.1560S} and known quasars now number in the hundreds of thousands \cite[see, e.g., the Million Quasar Catalogue, v7.5, hereafter Milliquas; ][]{2021arXiv210512985F}. Despite the proliferation of wide-area surveys across a range of wavelengths over the intervening fifty years, the search for bright quasars remains unfinished.

Here, we report on a spectroscopic investigation of a bright, blue, point-like source selected from the SkyMapper Southern Survey Data Release 2 \cite[SMSS DR2;][]{2018PASA...35...10W,2019PASA...36...33O}, which demonstrates that SMSS~J114447.77-430859.3 (SMSS~DR2 object\_id\footnote{The name and object\_id remain the same in SMSS DR3.} 84280208; hereafter, J1144) is a $z=0.83$ quasar. Aside from one blazar object at $z=0.6$ (PKS~1424+240), this makes J1144 the optically brightest quasar known above a redshift of 0.4.

Spectroscopy of J1144 was first acquired during a search for symbiotic binaries, in which cool giant stars accrete onto smaller companions, using SMSS DR2 \citep{2021PhDT........17L}. While any known active galactic nucleus \cite[AGN, as identified in SIMBAD;][]{2000A&AS..143....9W} was excluded, J1144 had only been identified as an AGN {\it candidate} by its near-IR and IR colours. It was identified as a candidate by \citet{2012ApJ...751...52E} using the Two Micron All-Sky Survey \cite[2MASS;][]{2006AJ....131.1163S} and the {\it Wide-field Infrared Survey Explorer} \cite[{\it WISE};][]{2010AJ....140.1868W,2011ApJ...731...53M}. Similarly, \citet{2015ApJS..221...12S} utilised the AllWISE\footnote{See \url{https://wise2.ipac.caltech.edu/docs/release/allwise/expsup/index.html}.} update to the IR dataset and selected J1144 as a quasar candidate from its IR colours alone. Notably, \citet{2019MNRAS.489.4741S} even estimated a photometric redshift for J1144 of $z=0.82$ from DR2 of the {\it Gaia} satellite mission \citep{2016A&A...595A...1G,2018A&A...616A...1G} and the unWISE \citep{2019ApJS..240...30S} revision to the photometry from {\it WISE}.

However, the {\it WISE} colours of symbiotic stars sometimes fall in typical AGN selection regimes, which is why such sources were not excluded from the symbiotic star search. In fact, the flickering, accretion-powered symbiotic star EF~Aql has an AGN-like $(W1-W2) \approx 0.9$ colour and was discovered via the UV-bright Quasar Survey \cite[UVQS;][]{2016AJ....152...25M,2016PASP..128b4201M,2017AN....338..680Z}. 
\citet{2021PhDT........17L} explored various SMSS selection mechanisms for different types of symbiotic stars, and J1144, being much bluer than most isolated cool giant stars, was caught by a $(u-g)$/$(u-v)$ colour-only selection. Thus, along with 232 other symbiotic star candidates, optical spectroscopy of J1144 was obtained with the goal of confirming the presence of a cool giant star with emission lines, indicative of symbiotic binarity. J1144 was the only observed source to show AGN emission lines.

In \S~\ref{sec:obs}, we describe the observations and data processing. In \S~\ref{sec:analysis}, we analyse the spectroscopic data and estimate the mass of the central black hole (BH). We summarise additional data available for J1144 in \S~\ref{sec:ancillary} and compare J1144 to other bright quasars in \S~\ref{sec:comparison}. \S~\ref{sec:discussion} discusses 
the outlook to further study and utilisation of J1144, as well as 
the prospects for additional such discoveries in the future.
Throughout the paper, we use Vega magnitudes for {\it Gaia} and IR data, and AB magnitudes for the SkyMapper passbands: $uvgriz$. 
We adopt a flat $\Lambda$CDM cosmology with $\Omega_{\rm m}=0.3$ and a Hubble-Lema\^itre constant of $H_0=70$~km~sec$^{-1}$~Mpc$^{-1}$.

\section{OBSERVATIONS AND DATA PROCESSING}
\label{sec:obs}

The spectroscopic portion of the symbiotic star program \cite[described in][]{2021PhDT........17L} principally used the South African Astronomical Observatory (SAAO) 1.9-meter telescope and its SpUpNIC instrument \cite[Spectrograph Upgrade: Newly Improved Cassegrain;][]{2019JATIS...5b4007C}. Following the initial classification of J1144 as an AGN, additional optical spectroscopic data was obtained at higher spectral resolution with the Australian National University (ANU) 2.3-meter telescope at Siding Spring Observatory (SSO) using the Wide Field Spectrograph \cite[WiFeS;][]{2007Ap&SS.310..255D,2010Ap&SS.327..245D}, and near-IR spectroscopy was obtained with the TripleSpec4.1 instrument \citep{2014SPIE.9147E..2HS} on the Southern Astrophysical Research (SOAR) 4.1-meter telescope.

\subsection{Optical spectroscopy with SAAO 1.9m / SpUpNIC}

SpUpNIC observations of J1144 were obtained on UT~2019-06-24 with the G7 grating. G7 is a low-resolution
grating, which covered $3300-8930$~\AA\ with a resolving power R$\sim$500. A BG38 filter was manually inserted into
the arc beam. The spectroscopic slit width was 2.24~arcsec in seeing of $\sim2$\arcsec, and a spatial binning of 2 pixels was used. The exposure time was 1200~s, with the object at an airmass of 1.2. 

Flux calibration was performed with observations of the spectrophotometric standard star, CD-32~9927 \citep{1994PASP..106..566H}\footnote{But see \citet{1999PASP..111.1426B} regarding correction of telluric features.}, obtained on the same night. Data reduction, including bias subtraction and flat-fielding, used the standard tasks in the Image Reduction and Analysis Facility \cite[IRAF;][]{1986SPIE..627..733T}. A second pass at flux scaling was performed by processing the spectrophotometric standard in the same way as the science spectra and determining the residual correction needed to align the flux with the model values. The final signal-to-noise ratio (S/N) of the SpUpNIC spectrum was $\sim50$ per pixel.

\subsection{Optical spectroscopy with ANU 2.3m / WiFeS}

WiFeS spectra were obtained on UT~2022-03-11 with two grating configurations. WiFeS is an integral field spectrograph and when used with a spatial binning of 2 pixels, it provides $1x1$~arcsec sampling over its $25x38$~arcsec field-of-view.

With the resolving power R$\sim$3000 gratings, an exposure of 600~s was obtained, covering the wavelength range $3250-9550$~\AA\ across the two cameras of the spectrograph (the RT560 beamsplitter was used). For the high-resolution gratings (R$\sim$7000), an exposure time of 900~s was used. The B7000 grating covered $4180-5540$~\AA, while the I7000 grating covered $6810-9040$~\AA. All observations were obtained near an airmass of 1.2 with seeing of $1.8-2$~arcsec in the $i$-band.

The spectrophotometric standard star, BD-12~2669 \citep{2010AAS...21546302H}, was observed immediately after the J1144 spectra. The raw frames were reduced with the {\sc Python}-based pipeline, PyWiFeS \citep{2014Ap&SS.349..617C}. We then extracted the spectra from the calibrated 3D data cubes using QFitsView\footnote{Available from \url{https://www.mpe.mpg.de/~ott/QFitsView/}}, selecting nearby source-free regions for sky subtraction. Variance and data-quality frames were extracted from the same regions. As with the SAAO data, a second iteration of flux scaling was performed by aligning the processed spectrophotometric standard spectrum to the model fluxes. The S/N in the final spectra ranged from 20-60 per pixel.

\begin{figure*}[ht]
\centering
\includegraphics[width=\textwidth]{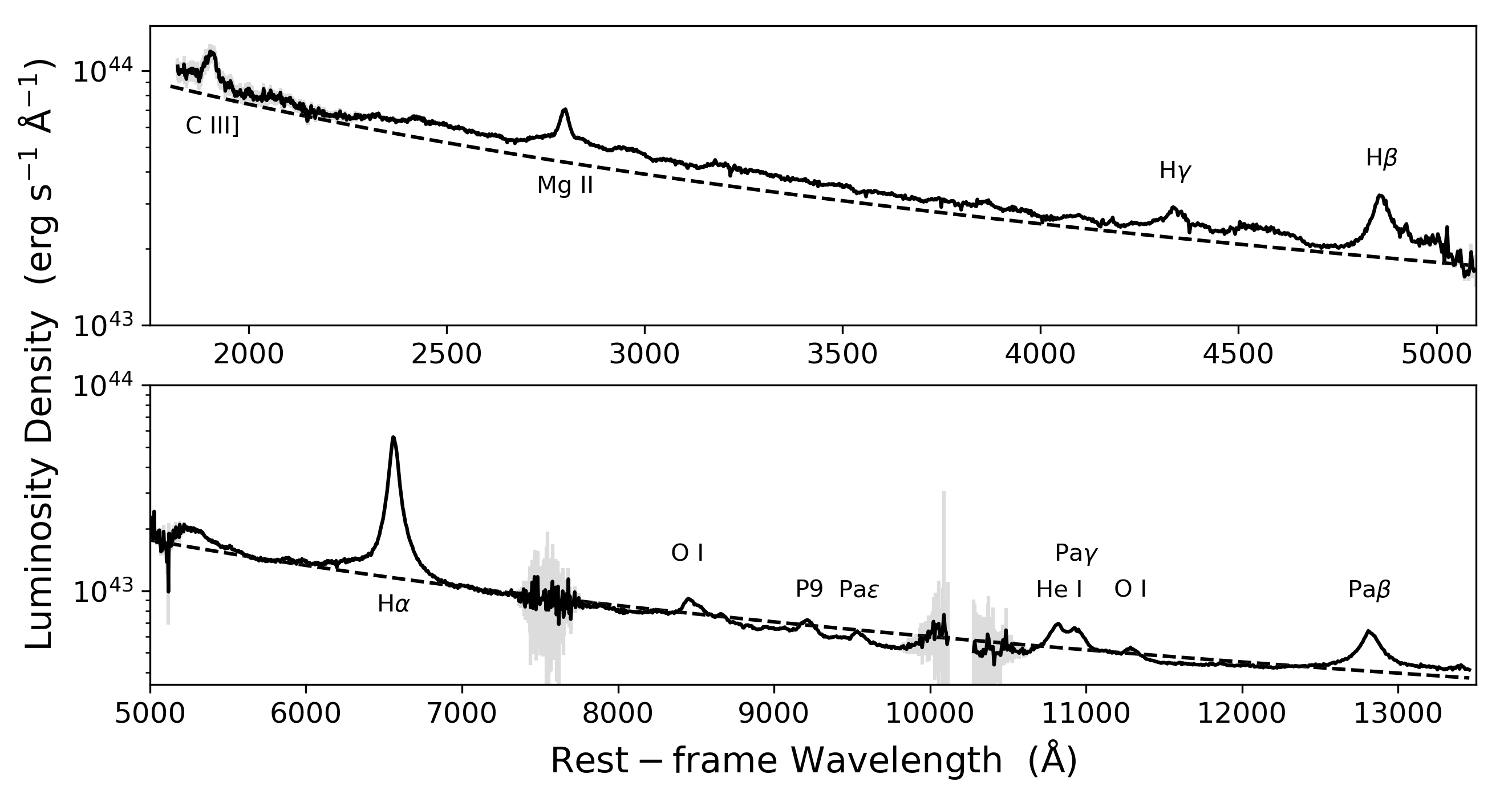}
\caption{Rest-frame spectrum of J1144, in units of erg~s$^{-1}$~\AA$^{-1}$. Uncertainties are shown with grey errorbars, typically smaller than the thickness of the line. The dashed line shows a power-law continuum with slope $\alpha_{\lambda}$=-1.56, which fits the spectrum well up to wavelengths of 7000~\AA. The spectrum shown here has been corrected for Galactic reddening, but not for any internal reddening.}
\label{fig:spectrum}
\end{figure*}

\subsection{Near-IR spectroscopy with SOAR / TripleSpec4.1}

We observed J1144 with TripleSpec4.1 on UT~2022-02-13 under NOIRLab program 2022A-389756 (PI: X.~Fan). TripleSpec4.1 utilises a fixed spectroscopic slit of $1.1\times28$~arcsec and produces cross-dispersed spectra that cover a simultaneous wavelength range from 0.95 to 2.47 microns, at a spectral resolution of $\sim3500$.

The observations were performed in three consecutive ABBA patterns, with 40s exposures at each dither position. The detector was read out with 4-pair Fowler sampling \citep{1990ApJ...353L..33F}. The seeing was 1~arcsec in $J$-band. Observations of the A0V star, HIP~56984, were obtained immediately prior to J1144 to serve as a telluric and flux standard.
The data were processed with the Spextool software package \citep{2004PASP..116..362C,2014ascl.soft04017C} written in the Interactive Data Language (IDL\footnote{See \url{https://www.l3harrisgeospatial.com/Software-Technology/IDL}.}), as modified for TripleSpec4.1\footnote{See \url{https://noirlab.edu/science/observing-noirlab/observing-ctio/observing-soar/data-reduction/triplespec-data}.}. Telluric correction was applied using the {\sc xtellcorr} package \citep{2003PASP..115..389V} in IDL. The final S/N was $50-100$ per pixel.

\section{Spectroscopic analysis}
\label{sec:analysis}

We normalise the spectra by anchoring them to the photometric data which best overlap the cleanest wavelength regions of each spectrum. This involves the $g$, $r$, and $i$ bands from SMSS DR3 and the 2MASS $H$-band for the near-IR spectrum. The bandpass details were retrieved from the Spanish Virtual Observatory (SVO) Filter Profile Service\footnote{See \url{http://svo2.cab.inta-csic.es/theory/fps/}.} \citep{2012ivoa.rept.1015R,2020sea..confE.182R}. There is good agreement amongst the calibrations provided by the available photometric bands (typically better than 5\%), consistent with the small levels of photometric variability discussed in Sec.~\ref{sec:ancillary}. 

We correct for Galactic reddening using the \citet{2019ApJ...886..108F} extinction curve, as implemented in the \linebreak {\sc dust\_extinction} Python package \citep{2021zndo...4658887G}. We assume $R_V=3.1$, and we take the $E(B-V)=0.123$~mag from \citet{1998ApJ...500..525S} and additionally apply the $\times0.86$ correction factor of \citet{2011ApJ...737..103S}.\footnote{The $E(B-V)$ map recommended by \citet{2021MNRAS.503.5351S}, from the generalised needlet internal linear combination (GNILC) analysis of the {\it Planck} 2015 data release \citet{2016A&A...596A.109P}, gives a consistent value of $0.121 \pm 0.004$~mag. \citet{2021MNRAS.503.5351S} also suggest retaining the rescaling factor of 0.86.} The spectra are combined as the weighted mean on a new wavelength grid which sampled the spectra in pixels at 200~km~s$^{-1}$ spacing. The strong emission lines in the J1144 spectrum (\ion{Mg}{ii}, H$\beta$, H$\alpha$, and Pa$\beta$) were used together (weighted mean) to provide a redshift estimate of $0.8314\pm0.0001$. The observed spectrum is then transformed to rest-frame wavelengths and to luminosity in units of erg~s$^{-1}$~\AA$^{-1}$, giving a velocity resolution of 109 km~s$^{-1}$~pixel$^{-1}$ and a final S/N between 100 and 250 per pixel. The combined spectrum is shown in Figure~\ref{fig:spectrum}.

When fitting the spectrum, we separately consider the wavelength regimes around \ion{C}{iii}], \ion{Mg}{ii}, H$\beta$, H$\alpha$, and Pa$\beta$. For each wavelength region, we first fit a combined power-law continuum and iron template. The best power-law slope for the combined UV/optical range is found to be $\alpha_{\lambda}=-1.56$, although a flatter slope of -0.79 is a better fit for the near-IR, likely reflecting the contributions of hot dust in the region longward of 1~$\mu$m (cf. the {\it WISE} photometry in Fig.~\ref{fig:sed}). Because of the impact of the iron model on the remaining emission line profiles, we test the systematic effects of adopting various iron emission templates in the UV and optical portions of the spectrum. For the UV iron templates, we use those of \citet[][; implementing a combination of \citet{2001ApJS..134....1V}, 
\citet{2006ApJ...650...57T}, and \citet{2007ApJ...662..131S}]{2012ApJ...753..125S}, \citet{2006ApJ...650...57T}, and \citet{2016MNRAS.460..187M}, while for the optical templates, we use those of \citeauthor{1992ApJS...80..109B}\citetext{1992; BG92, hereafter}, \citet{2006ApJ...650...57T}, \citet{2008ApJ...675...83B}, and \citet{2022ApJS..258...38P}. The velocity broadening of the template is a free parameter in each fit. On average, the best-fit iron full width at half-maximum (FWHM) was $\sim2500$~km~s$^{-1}$. The systematic errors arising from the iron template choice are square-added to the statistical errors in the fit results below.

For the subsequent steps of the spectral modelling, 
the emission lines are each fit with a sum of three Gaussian profiles. 
We estimate the statistical uncertainties on the fit parameters via a Monte Carlo approach, taking the RMS from 50 realisations in which the flux at each pixel is varied according to the error spectrum.
Because of the weak [\ion{O}{iii}] and [\ion{S}{ii}] emission and lack of evident narrow-line contribution to the Balmer lines, even in the original $R\sim7000$ spectrum of H$\beta$, no narrow-line subtraction is performed for the Balmer lines or the [\ion{N}{ii}] lines near H$\alpha$. Integrating $3\times$ the error spectrum over spectral windows of $\pm200$~km~s$^{-1}$ around [\ion{O}{iii}] 5007~\AA\ and [\ion{S}{ii}] $6716+6731$~\AA\ provides conservative upper limits of $5\times 10^{42}$ and $3\times 10^{42}$~erg~s$^{-1}$, respectively; however, the blending of H$\alpha$ with [\ion{N}{ii}] precludes a similar upper limit estimate for the latter.

\begin{figure*}[t]
\centering
\includegraphics[width=\textwidth]{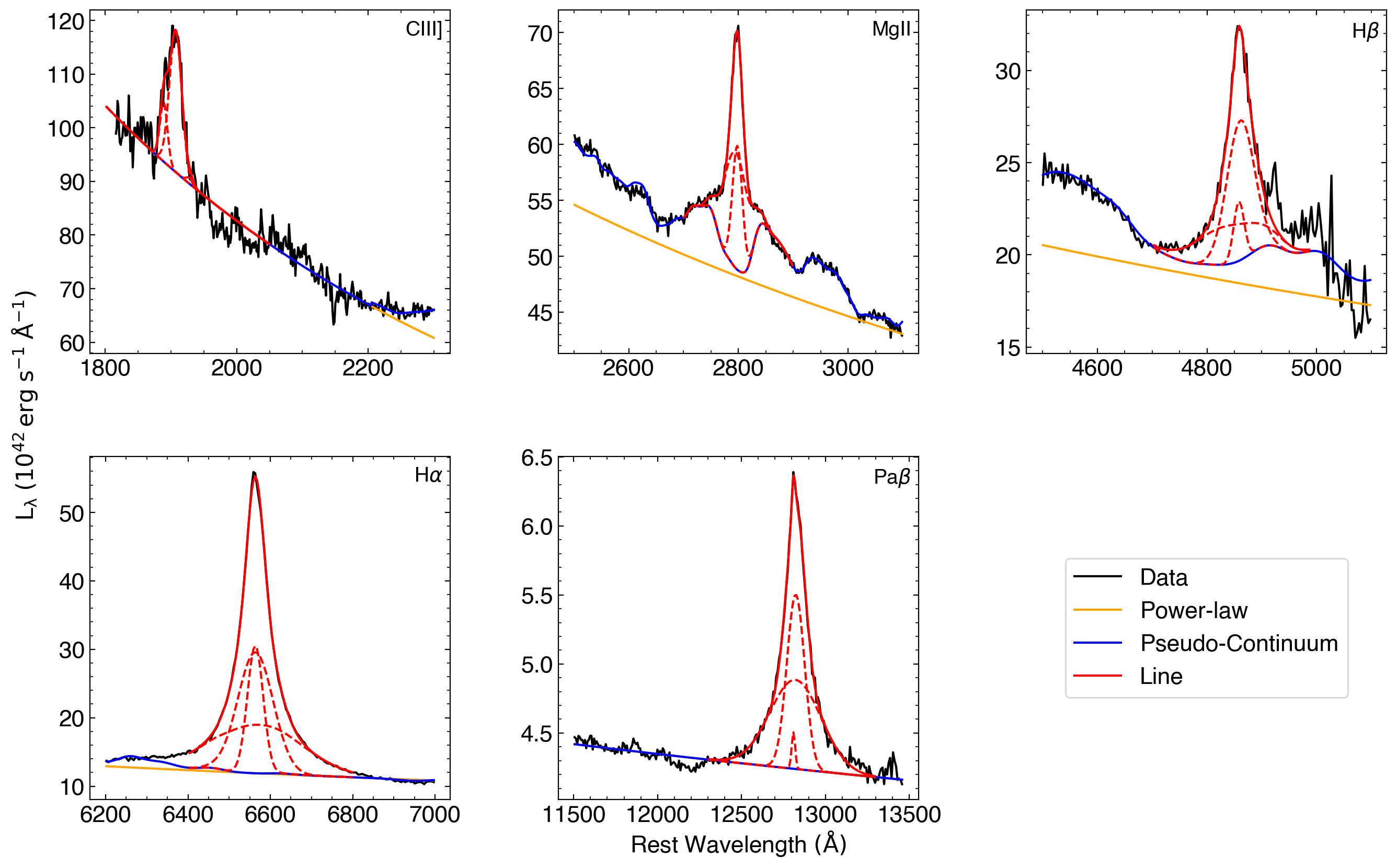}
\caption{Emission line fits to \ion{C}{iii}], \ion{Mg}{ii}, H$\beta$, H$\alpha$, and Pa$\beta$, as indicated in each panel. Black lines indicate the data. The model is plotted with progressively added elements: the power-law continuum (orange), then the pseudo-continuum from the broadened iron template (blue), then the emission line fits (red). The red dashed lines indicate the three Gaussian profiles used to fit each line. The particular fits shown here use the \citet{2012ApJ...753..125S} and BG92 templates in the UV and optical, respectively.}
\label{fig:fits}
\end{figure*}

The emission line fits are shown in Figure~\ref{fig:fits}. The \ion{C}{iii}] fit is poorly constrained, because of limited wavelength coverage, the lower spectral resolution of the SAAO data from which it is principally observed, and lack of de-blending with \ion{Al}{iii} and \ion{Si}{iii}]. As a result, the parameters are omitted from Table~\ref{tab:lines}. The features redward of H$\beta$ are likely iron lines that have not been well modelled because of errors in the flux calibration at the long-wavelength limit of the optical spectra.

From the sum of the Gaussian fits, we determine the velocity shift (derived from the peak of the line profile), the integrated line luminosity, the FWHM, and the second moment of the line profile ($\sigma_{\rm line}$). The emission line properties are summarised in Table~\ref{tab:lines}. As was the case for the continuum luminosities, the emission line luminosity errors are dominated by the photometric calibration uncertainties.

\begin{table}[h]
\begin{threeparttable}
\caption{Emission Line Fit Results}\label{tab:lines}
\begin{tabular}{lcccc} \toprule
Line & Line Shift\tnote{a} & Luminosity & FWHM & $\sigma_{\rm line}$ \\
 & (km~s$^{-1}$) & (10$^{44}$~erg~s$^{-1}$) & (km~s$^{-1}$) & (km~s$^{-1}$) \\ \midrule
\ion{Mg}{ii} & $-180 \pm 110$ & $6.6 \pm 0.9$ & $3100 \pm 240$ & $1670 \pm 240$ \\
H$\beta$ & $-120 \pm 70$ & $9.0 \pm 1.0$ & $3200 \pm 240$ & $2750 \pm 310$ \\
H$\alpha$ & $20 \pm 10$ & $46.8 \pm 0.8$ & $3390 \pm 40$ & $3380 \pm 170$ \\
Pa$\beta$ & $-10 \pm 100$ & $4.2 \pm 0.1$ & $3450 \pm 60$ & $2890 \pm 80$ \\ \bottomrule
\end{tabular}
\begin{tablenotes}
\item[a] Measured with the emission line peak.
\end{tablenotes}
\end{threeparttable}
\end{table}

\subsection{Luminosity of J1144}\label{sec:lum}

From the power-law continuum fits, we determine the luminosities at several rest-frame wavelengths of interest:\\
$\log_{10}$ ( $\lambda$\,L$_{\lambda}$(3000\AA) / erg~s$^{-1}$ ) = 47.12 $\pm$ 0.04,\\
$\log_{10}$ ( $\lambda$\,L$_{\lambda}$(5100\AA) / erg~s$^{-1}$ ) = 46.94 $\pm$ 0.04,\ and\\
$\log_{10}$ ( $\lambda$\,L$_{\lambda}$(1$\mu$m) / erg~s$^{-1}$ )~~~=~46.67 $\pm$ 0.04,
\\
where the errors are dominated by the 0.1~mag uncertainties in the flux calibration but do not incorporate the small additional factor of source variability (see Sec.~\ref{sec:ancillary}). The luminosity at 3000~\AA\ translates into an absolute magnitude of $M_{300\rm nm} = -28.70$~mag (AB). To aid comparison with higher-redshift samples, we extrapolate the best-fit continuum power-law (with $\alpha_{\lambda}=-1.56$) to shorter wavelengths and find $\log_{10}$ ( $\lambda$\,L$_{\lambda}$(1450\AA) / erg~s$^{-1}$ ) = 47.2 or $M_{145\rm nm} = -28.36$~mag (AB). Using the prescription of \citet{2006AJ....131.2766R}, we find $M_{i}(z=2) = -29.74$~mag (AB).

We adopt the bolometric correction (BC) for 3000~\AA\ from \citet{2012MNRAS.422..478R,2012MNRAS.427.1800R}, although their BC values for 5100~\AA\ (with spectral slope correction) or the BC values from  \citet{2019MNRAS.488.5185N} yield similar results. The bolometric luminosity of J1144 is $(4.7\pm1.0)\times10^{47}$~erg~s$^{-1}$, where the uncertainty accounts for the variation arising from different BC assumptions. For the canonical radiative efficiency of 0.1 \cite[e.g.,][]{2002MNRAS.335..965Y}, this equates to an accretion rate of $\sim80$~M$_{\odot}$~yr$^{-1}$.

\subsection{BH Mass of J1144}\label{sec:mbh}

The wide wavelength coverage of our spectroscopic data provides access to several emission lines from the broad line region (BLR), which can be used to estimate the mass of the central BH in J1144 through the virial relations. Anchored to the H$\beta$ reverberation mapping results from the local AGN sample \cite[see][]{2004ApJ...613..682P}, the virial relations rely on a single epoch of line width and luminosity measurements to infer the velocity of the BLR gas around the BH, as well as the characteristic distance from the BH to the line-emitting gas in question \cite[see, e.g.,][and references therein]{2021iSci...24j2557C}. Each emission line may have its own velocity and distance, which should yield consistent mass estimates under the assumption that the gravity of the BH dominates the gas dynamics. However, we do note that J1144 represents an extrapolation of a factor of $\sim10$ in luminosity compared to the well measured H$\beta$ reverberation mapping sample of \citet{2013ApJ...767..149B}.

We adopt the $M_{\rm BH}$ relation parameters indicated in Table~\ref{tab:virial} for a functional form of 
\begin{equation}
    M_{\rm BH} = 10^{A} \times (FWHM / 10^{3})^B \times (L_{\rm cont/line} / 10^{44})^{C}
\end{equation}
for the emission line FWHM in units of km~s$^{-1}$, and the luminosity of either continuum or emission line in units of erg~s$^{-1}$. With the relations being primarily drawn from \citet{2020ApJ...901...35L}, the $A$ parameters in Table~\ref{tab:virial} have been renormalised to that paper's adopted virial factor of $f=1.12$ \cite[from][]{2015ApJ...801...38W} for these FWHM-based measurements. 

\begin{table}
\begin{threeparttable}
\caption{Virial Relations}\label{tab:virial}
\begin{tabular}{lccccc} \toprule
Emission & Luminosity & A & B & C & Reference\\ 
Line & (cont/line) & & & & \\ \midrule
\ion{Mg}{ii} & 3000~\AA & 7.04 & 2.0 & 0.5 & 1 \\ 
H$\beta$ & 5100~\AA & 6.87 & 2.0 & 0.533 & 1 \\ 
H$\beta$ & H$\beta$ & 7.79 & 2.0 & 0.54 & 1 \\ 
H$\alpha$ & H$\alpha$ & 7.51 & 2.06 & 0.46 & 2 \\
Pa$\beta$ & 1~$\mu$m & 7.15 & 1.76 & 0.44 & 3\tnote{a} \\ \bottomrule
\end{tabular}
\begin{tablenotes}[flushleft]
\item References: 1 - \citet{2020ApJ...901...35L}; 2 - \citet{2015ApJ...801...38W}; 3 - \citet{2013MNRAS.432..113L}.
\item[a] Rescaled from $f=1.4$ to $f=1.12$.
\end{tablenotes}
\end{threeparttable}
\end{table}

With the emission line measurements of Table~\ref{tab:lines} and the continuum luminosities indicated above, we derive several complementary BH mass estimates. The $M_{\rm BH}$ values we estimate are presented in Table~\ref{tab:mass}. For \ion{Mg}{ii}, we do not apply the mass correction factors from \citet{2020ApJ...901...35L} for the emission line shape (FWHM/$\sigma_{\rm line}$ ratio) and spectral slope, which would increase that mass estimate by 0.3~dex and make it more discrepant with the other emission line results. The different emission lines produce BH mass estimates that span the range from $(1.9-3.8)\times 10^{9}$~M$_{\odot}$. The high S/N of our spectroscopic data mean the statistical errors on the BH mass estimate are small compared to the systematic errors. \citet{2020ApJ...903..112D} estimate the intrinsic scatter to be 0.371~dex for the best-measured FWHM-based virial relation (using H$\beta$ and 5100~\AA), and the systematic errors for the other estimates are likely to be larger. Thus, we take the mean value of our five measurements as the best estimate for the BH mass in J1144, $M_{\rm BH} = 2.6 \times 10^{9}$~M$_{\odot}$, and conservatively adopt an uncertainty of 0.5~dex \cite[cf.][]{2009ApJ...699..800V}.
Our measurements of the bolometric luminosity and BH mass yield an Eddington ratio of $\approx 1.4$ for J1144.

\begin{table}
\begin{threeparttable}
\caption{BH Mass Estimates}\label{tab:mass}
\begin{tabular}{lcccc} \toprule
Emission Line & Luminosity & BH Mass $\pm$ Stat. Error\\
 & & (10$^{9}$~M$_{\odot}$) \\ \midrule
\ion{Mg}{ii} & 3000~\AA & $3.83 \pm 0.62$ \\
H$\beta$ & 5100~\AA & $2.81 \pm 0.44$ \\
H$\beta$ & H$\beta$ & $2.07 \pm 0.34$ \\
H$\alpha$ & H$\alpha$ & $2.35 \pm 0.06$ \\
Pa$\beta$ & 1~$\mu$m & $1.87 \pm 0.09$ \\\midrule
\multicolumn{2}{l}{{\bf Combined Estimate $\pm$ Sys. Error}} & {\bf 2.6 $^{+5.6}_{-1.8}$} \\\bottomrule
\end{tabular}
\end{threeparttable}
\end{table}

In Table~\ref{tab:summary}, we present a summary of the observed and derived properties of J1144.

\begin{table}[t!]
\begin{threeparttable}
\caption{Summary of J1144 Properties}\label{tab:summary}
\begin{tabular}{llll} \toprule
Property & Value & Unit & Notes\\ \midrule
$\alpha$ (J2000) & $176.199041$ & deg & SMSS DR3 \\
$\delta$ (J2000) & $-43.149829$ & deg & SMSS DR3 \\
$l$ & $290.222$ & deg & SMSS DR3 \\
$b$ & $+18.071$ & deg & SMSS DR3 \\
E(B-V) & $0.123$ & mag & [1] \\ 
SMSS {\tt object\_id} & 84280208 & \ldots & DR3 \\
SMSS $u$ & $14.974\pm0.029$ & mag (AB) & DR3 \\
SMSS $v$ & $15.026\pm0.033$ & mag (AB) & DR3 \\
SMSS $g$ & $14.534\pm0.014$ & mag (AB) & DR3 \\
SMSS $r$ & $14.424\pm0.017$ & mag (AB) & DR3 \\
SMSS $i$ & $14.270\pm0.007$ & mag (AB) & DR3 \\
SMSS $z$ & $14.097\pm0.006$ & mag (AB) & DR3 \\
{\it Gaia} {\tt source\_id} & \multicolumn{2}{l}{5379240246670899584} & DR3 \\
{\it Gaia} $G$ & $14.3887\pm0.0028$ & mag (Vega) & DR3 \\
{\it Gaia} $B_{P}$ & $14.6397\pm0.0036$ & mag (Vega) & DR3 \\
{\it Gaia} $R_{P}$ & $13.9321\pm0.0040$ & mag (Vega) & DR3 \\
2MASS $J$ & $12.806\pm0.024$ & mag (Vega) & \\
2MASS $H$ & $12.563\pm0.022$ & mag (Vega) & \\
2MASS $K_{s}$ & $11.877\pm0.024$ & mag (Vega) & \\
{\it WISE} $W1$ & $10.272\pm0.006$ & mag (Vega) & AllWISE \\
{\it WISE} $W2$ & $9.103\pm0.006$ & mag (Vega) & AllWISE \\
{\it WISE} $W3$ & $6.741\pm0.007$ & mag (Vega) & AllWISE \\
{\it WISE} $W4$ & $4.705\pm0.018$ & mag (Vega) & AllWISE \\
\\ \midrule \multicolumn{2}{l}{Derived Quantities} \\ \midrule
Redshift & $0.8314\pm0.0001$ & \ldots & \\
$M_{\rm BH}$ & $2.6\times 10^{9}$ & M$_{\odot}$ & \\
$\lambda$\,L$_{\lambda}$(3000\AA) & $1.3\times 10^{47}$ & erg~s$^{-1}$ & \\
$\lambda$\,L$_{\lambda}$(5100\AA) & $8.7\times 10^{46}$ & erg~s$^{-1}$ & \\
$\lambda$\,L$_{\lambda}$(1$\mu$m) & $4.7\times 10^{46}$ & erg~s$^{-1}$ & \\
$L_{\rm bol}$ & $4.7\times 10^{47}$ & erg~s$^{-1}$ & \\
$M_{300nm}$ & $-28.70$ & mag (AB) & \\
$M_{145nm}$ & $-28.36$ & mag (AB) & \\
$M_{i}(z=2)$ & $-29.74$ & mag (AB) & \\
Eddington Ratio & $1.4$ & \ldots & \\
\bottomrule
\end{tabular}
\end{threeparttable}
\begin{tablenotes}
\item[\protect{[1]}] From \citet{1998ApJ...500..525S}.
\end{tablenotes}
\end{table}

\subsection{Continuum Slope and Internal Reddening}

Typical thin disk \citep{1973A&A....24..337S} and slim disk \citep{1988ApJ...332..646A} models of BH accretion predict UV/optical continuum emission having a power-law slope of $\alpha_{\lambda} \approx -2.3$ ($\alpha_{\nu} \approx +0.3$), though real quasars are rarely observed to be so blue \citep{2016ApJ...824...38X}. Taking such a spectral slope as the limiting case, we can assess the maximum amount of internal reddening that may be present in the emitted spectrum of J1144.

For a UV-flat reddening curve like that of \citet[][GB07, hereafter]{2007arXiv0711.1013G}, we find that a maximum intrinsic E(B-V) of 0.17~mag provides a reasonable fit to the data. In contrast, a UV-steep reddening curve -- one lacking the strong bump at 2175~\AA\ -- such as the \citet{2003ApJ...594..279G} model for the star-forming bar of the Small Magellanic Cloud, implies a maximum E(B-V) of 0.10~mag to avoid over-correcting \ion{C}{iii}], but then under-corrects the spectrum near \ion{Mg}{ii}. (Reddening curves retaining the 2175~\AA\ bump perform even worse in the regime between \ion{C}{iii}] and \ion{Mg}{ii}.) The GB07 reddening correction would lift the 3000~\AA\ luminosity by 0.34~dex, which might be expected to increase each of the BH mass and the Eddington ratio by roughly half that margin. However, because the sources anchoring the virial relations have not been corrected for internal reddening, the appropriate adjustments for J1144 would be reduced in amplitude.

It is also worth highlighting that the power-law prescription for the rest-frame UV spectrum breaks down in accretion disk models at higher BH masses, as the high-energy turnover migrates to longer wavelengths \cite[e.g., see][]{2018A&A...612A..59C}. For $M_{\rm BH} \sim 10^{9}$~M$_{\odot}$, the departure from a power-law exceeds 0.1~magnitudes between 2000 and 3000~\AA, suggesting extreme care must be taken to disentangle the intrinsic spectral shape from any internal reddening. Observed-frame UV spectroscopy of J1144 should contribute to our ability to remove such degeneracies.

Finally, reddening will lead to an enhancement of the Balmer decrement, i.e., the H$\alpha$/H$\beta$ flux ratio. Low-redshift quasars with blue spectral slopes (implying little intrinsic reddening) are often found to have Balmer decrements of 3.1 \citep{2008MNRAS.383..581D}. Our observed Balmer decrement of 5.1 would imply E(B-V) $\approx 0.4$~mag, depending on the reddening curve adopted. However, \citet{2017MNRAS.467..226G} has argued that quasar Balmer decrements are consistent with an assumption of intrinsic Case B recombination and a flux ratio of 2.72, which would suggest even more reddening in J1144: E(B-V) $\approx 0.5$~mag. In either case, such high reddening appears to be incompatible with the expected spectral slopes, suggesting an intrinsically higher Balmer decrement, which can arise from optical depth effects redistributing H$\beta$ photons into H$\alpha$ and Pa$\alpha$ \citep{1960ApJ...131..202P,1975MNRAS.171..395N}.

\section{Ancillary Datasets}
\label{sec:ancillary}

The unusually bright nature of J1144 raises the question of whether it has historically been fainter, contributing to its long-standing anonymity. 

\subsection{Optical data}

On UT 1890-05-26, the 8-inch "Bache doublet" telescope at "Mount Harvard" near Chosica, Peru, obtained a 60-minute exposure showing J1144. The photographic glass plate (b5269) has been digitised, and astrometrically and photometrically calibrated by the Digital Access to a Sky Century @ Harvard (DASCH) project\footnote{See \url{https://library.cfa.harvard.edu/dasch}.} \citep{2008arXiv0811.2005L,2013PASP..125..857T}. The brightness of J1144 in the 1890 image is estimated to be $14.80\pm0.16$~mag (AB), calibrated to Pan-STARRS $g$-band using the Asteroid Terrestrial impact Last Alert System (ATLAS) All-Sky Stellar Reference Catalog \citep{2018ApJ...867..105T}. Similar plates available through DASCH from more recent epochs show J1144 with estimated $g$-band magnitudes between 13.9 and 15.1~mag (AB; omitting highly uncertain measurements or those close to the plate's limiting depth). The heterogeneity of the data precludes a more detailed analysis, but the DASCH photographic archives indicate that the optical brightness of J1144 has not varied by more than a factor of $\sim2$ in the last 130 years.

Considering data focused on the blue end of the optical spectrum, photographic glass plates taken by the UK Schmidt telescope at SSO on UT 1977-03-21, as part of the ESO/SERC Southern Sky Survey, measured $B_{J}=14.7\pm0.4$~mag (Vega), as catalogued\footnote{See \url{http://www-wfau.roe.ac.uk/sss/index.html}.} by the SuperCOSMOS Sky Survey \citep{2001MNRAS.326.1279H,2001MNRAS.326.1295H}.
The $B_{J}$ and SMSS $g$ bandpasses are similar
\cite[e.g., see the $B_{J}$ throughput compared to the Sloan Digital Sky Survey $g_{\rm SDSS}$ in][]{2005MNRAS.360..839R} and the ($B_{J}-g$) colours of 3673 quasars in the redshift range $0.6-1.0$ from the 2dF and 6dF QSO Redshift Surveys \cite[2QZ/6QZ;][]{2004MNRAS.349.1397C} show a median of 0.10~mag with a scaled median absolute deviation (SMAD) of 0.38~mag. With an SMSS DR3 $g$-band magnitude of $14.534\pm0.014$~mag (AB), we conclude there has been no significant variation in the J1144 brightness at rest-frame $\sim2700$~\AA\ compared to 45 years ago.

On more recent timescales, the SMSS DR3 dataset incorporates images of J1144 acquired between February 2015 and June 2018. Across that time window, each of the six SMSS filters shows a brightening of $\approx0.1$~mag, with $7-9$ epochs per filter. Denser time sampling is available from the ATLAS telescopes \citep{2018PASP..130f4505T,2020PASP..132h5002S}, which have observed J1144 at a high rate (typically over 200 times per year) since December 2017\footnote{See \url{https://fallingstar-data.com/forcedphot/}.}. Both the $o$ ("orange"; comprising 80\% of the data) and $c$ ("cyan") filters show a brightening of 0.2~mag relative to the earliest ATLAS epoch (January 2016), with a peak roughly in May 2020. Together, these datasets provide a consistent picture of modest brightness variations over timescales of days to years.

To probe even shorter timescales, we examined the J1144 data available from two $\sim$month-long visits by the {\it Transiting Exoplanet Survey Satellite} \cite[{\it TESS};][]{2015JATIS...1a4003R}, one beginning on UT 2019-03-26 (during the primary mission; Sector~10) and one beginning on 2021-04-02 (first extended mission; Sector~37). In the {\it TESS} Input Catalog \cite[TIC;][]{2022yCat.4039....0P}, J1144 has the designation TIC~61537875, but it was not selected for the Candidate Target List. As a result, photometry is only available from the Full Frame Image (FFI) dataset, which were acquired every 30~minutes in the primary mission and every 10~minutes in the first extended mission. The coarse spatial resolution of {\it TESS} results in J1144 being heavily blended with TIC~61537878, a star 30~arcsec away that is $\sim1$~mag brighter in {\it TESS}'s wide bandpass ($600-1000$~nm). As a result, precise photometry is difficult to obtain, but we note no significant fluctuations above 1\% in the combined light curve of the quasar and star.

\subsection{IR data}

\begin{figure}[t]
\centering
\includegraphics[width=\textwidth]{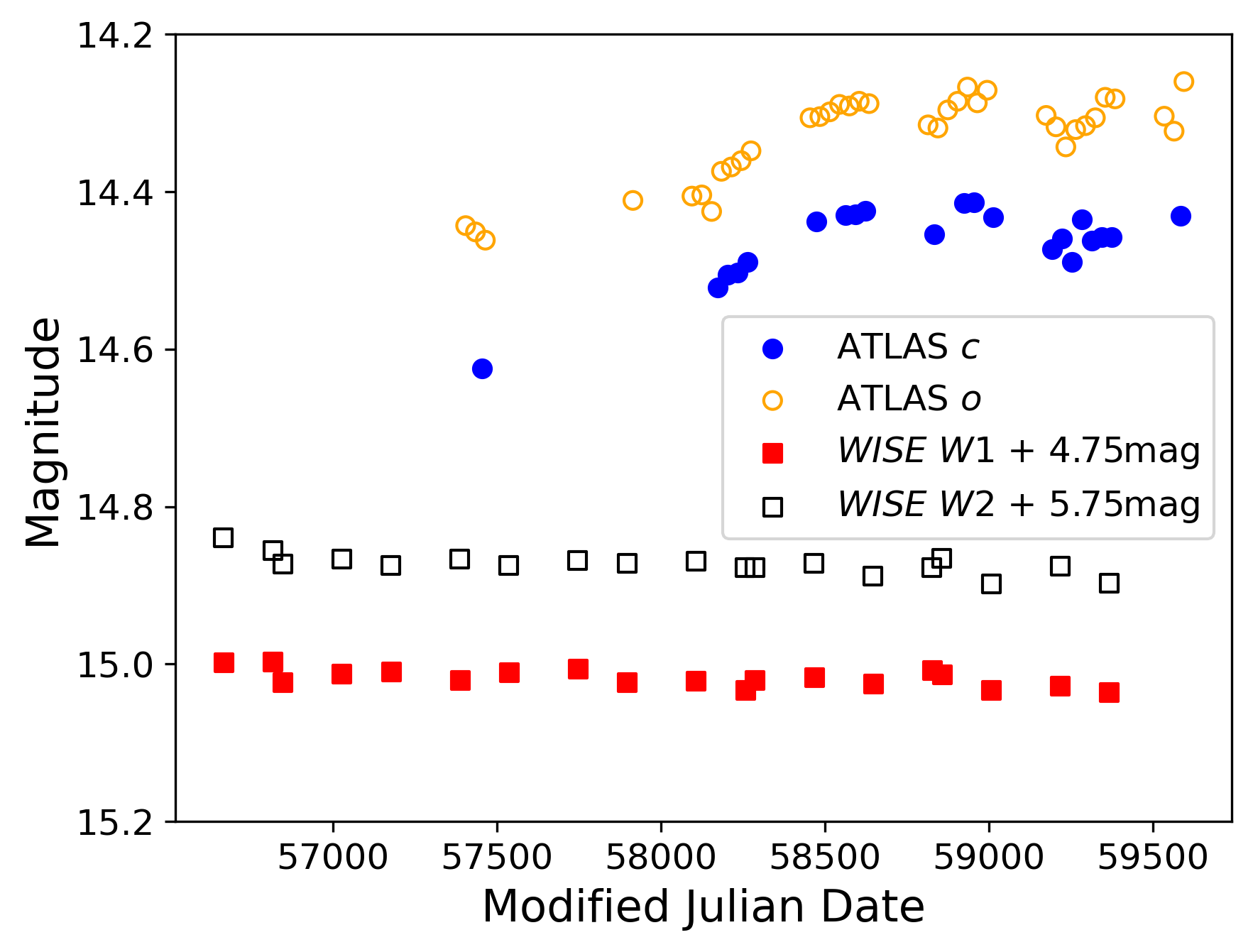}
\caption{ATLAS and {\it WISE} light curves for J1144 over the past 8 years. Photometry was binned to 30-day median values for each bandpass. The ATLAS photometry is in AB magnitudes, while the the IR photometry is in Vega magnitudes and has been shifted vertically for convenience. Any increase in the optical brightness would take a decade to be reflected in the dust luminosity because of the large dust sublimation region around luminous quasars like J1144.}
\label{fig:lc}
\end{figure}

Turning to the IR, the NEOWISE 2022 Data Release\footnote{See \url{https://wise2.ipac.caltech.edu/docs/release/neowise/}.} \citep{2014ApJ...792...30M} provides $W1$ and $W2$ photometry for J1144 at more than 250 epochs between UT 2014-01-09 and 2021-06-19. At rest-frame equivalents of 1.8 and 2.5~$\mu$m for $W1$ and $W2$, respectively, the {\it WISE} data is dominated by the hot dust near the quasar, rather than the quasar's accretion disk that is probed at shorter wavelengths. In Figure~\ref{fig:lc}, we show the {\it WISE} and ATLAS light curves, binned to 30-day median values. 

In contrast to the ATLAS photometry, J1144 exhibits a very slight fading in both IR bands over that time period, with an amplitude of $\sim0.05$~mag, comparable to the level of intraday photometric scatter. This uncorrelated behaviour can be understood in the context of the multi-year time lags for dust reverberation that would be expected from the large dust sublimation radius implied by the high luminosity of J1144. For the luminosity of J1144 ($1.2 \times 10^{14}$~L$_{\odot}$), the predicted time lag for the IR response to optical variations would be 9.7~years \citep{2019ApJ...886...33L}, slightly longer than the time span shown in Figure~\ref{fig:lc}. Thus, the steadiness of the IR photometry since the start of {\it WISE} operations implies no significant and lengthy changes in the quasar luminosity over the past $\sim$20 years. 

As the recent increase in luminosity shown by the ATLAS light curves propagates into the dust surrounding the quasar, we could expect to see the IR similarly brighten within the next decade. However, the relative amplitude of IR variability in response to optical fluctuations is extremely broad, between factors of 0.1 and 10 for high-luminosity, high-time-lag sources \cite[][cf. their Fig.~14]{2019ApJ...886...33L}. Future monitoring by {\it WISE} and the Dynamic REd All-sky Monitoring Survey \cite[DREAMS;][]{2020SPIE11203E..07S} may reveal the amplitude of the dust response in J1144.

\subsection{X-ray data}

In X-rays, there are no point-source counterparts to J1144 catalogued in the Second {\it ROSAT} all-sky X-ray Survey \cite[2RXS;][]{2016A&A...588A.103B}. Cross-matching confirmed quasars from Milliquas, approximately 75\% of the {\it Gaia} $G<16$~mag (Vega) quasars in the redshift range $z=0.7-0.9$ have 2RXS counterparts. For our extinction-corrected 2500~\AA\ flux of $F_{\nu} = 3.8\times10^{-26}$ erg~s$^{-1}$~cm$^{-2}$~Hz$^{-1}$, we would expect a 2~keV X-ray flux of  $\sim2\times10^{-30}$ erg~s$^{-1}$~cm$^{-2}$~Hz$^{-1}$ \citep{2021A&A...655A.109B}, roughly a factor of 10 greater than the nominal 2RXS flux limit \citep{2016A&A...588A.103B}. Whether the non-detection is an indication of intrinsic X-ray weakness or absorption (local to the quasar or intervening) may be clarified by forthcoming data releases from {\it SRG/eROSITA} \cite[the {\it Spectrum-Roentgen-Gamma} satellite's {\it extended ROentgen Survey with an Imaging Telescope Array};][]{2021A&A...647A...1P}.

\subsection{Radio data}

The closest radio detection in DR1 of the Rapid ASKAP Continuum Survey\footnote{See the interactive Hierarchical Image Survey (HiPS) map in the Aladin sky atlas \citep{2000A&AS..143...33B} under "Collections-Image-Radio-RACS".} \cite[RACS;][]{2020PASA...37...48M,2021PASA...38...58H} is 48~arcsec from J1144, which at $z=0.83$ corresponds to more than 350~kpc projected distance. If we assume an association between J1144 and RACS-DR1~J114451.8-430920, then the flux density of $5.1\pm0.5$~mJy at 887.5~MHz implies a rest-frame 5~GHz luminosity density of $7.6\times10^{31}$~erg~s$^{-1}$~Hz$^{-1}$ (for a spectral index of $\nu^{-0.6}$). For an optical (rest-frame 4400~\AA) luminosity density of $1.4\times10^{32}$~erg~s$^{-1}$~Hz$^{-1}$ (derived from the global power-law continuum shown in Figure~\ref{fig:spectrum}), this gives an upper limit to the radio-loudness \cite[using the definition of][]{1989AJ.....98.1195K} of $R < 0.54$. Thus, even a putative association\footnote{We further caution the reader that the visual appearance of this RACS source is plausibly of a core and (low-significance) two-lobe structure unassociated with J1144, but without any detected counterpart in SMSS DR3, VHS, or unWISE. Deeper $i$- and $z$-band images from the DECam instrument \citep{2015AJ....150..150F} on the CTIO 4m telescope, taken as part of programs 2017A-0260 (PI: M.\ Soares-Santos) and 2019A-0272 (PI: A.\ Zenteno), do not reveal any sources aligned with the centre of the radio "core".} between the RACS DR1 source and J1144 leaves the quasar in the radio-quiet regime.

The absence of strong X-ray and radio emission, in conjunction with the low levels of UV-to-IR variability, make it unlikely for J1144 to have a relativistically beamed jet. Thus, we conclude that J1144 is not a blazar.

\subsection{On the possibility of gravitational lensing} 

Given the high luminosity, it is natural to wonder if the source is gravitationally lensed \cite[e.g.,][]{2019ApJ...870L..11F}. To check for a small-separation galaxy lens, we examine the corrected $B_{P}$ and $R_{P}$ flux excess, $C^{*}$, from {\it Gaia} EDR3 \citep{2021A&A...649A...3R}, which makes a standardised comparison between the flux measured in the 0.35-arcsec-wide $G$-band photometric aperture with the integrated fluxes measured in the 3.5-arcsec-wide apertures for the $B_{P}$ and $R_{P}$ photometry (integrating along the wavelength dimension of the low-resolution spectra). With $C^{*} = 0.023$, the variation in flux measured by the different extraction apertures for the {\it Gaia} photometry is within 2$\sigma$ of the 0-value expected for point sources of the brightness of J1144 (where $\sigma = 0.012$). Amongst the small-separation gravitationally lensed quasars\footnote{We utilise the catalogue at \url{https://research.ast.cam.ac.uk/lensedquasars/} compiled by C.~Lemon.} that are associated with a single {\it Gaia} EDR3 source, Q1208+101 has the smallest $C^{*}$ value at 0.168, a factor of 7 times larger than J1144. Similarly, the $G$-band variability proxy of \citet{2021A&A...648A..44M} has a value of 0.015, suggesting that the {\it Gaia} small-aperture measurements across a range of scan directions have found peak-to-peak flux variations of $\sim0.05$~mag (cf. their Sect.~3). 

While we conclude that there is no indication of gravitational lensing for J1144 in the existing {\it Gaia} data, the lack of microlensing-induced flux variations evident in the recent photometric sampling described above cannot fully exclude the existence of a lensing galaxy, as \citet{2011ApJ...738...96M} found that roughly half of lensed quasars are likely to be in a "demagnified valley" in any given 10-year period. Thus, a high-spatial-resolution imaging study of J1144 would be of great interest.

\section{Comparison with other bright quasars}
\label{sec:comparison}

As the single brightest quasar in the sky, 3C~273 is an important benchmark for luminous quasars, as its long observational history and low redshift have made it a forefront laboratory for exploring accretion processes \cite[e.g.,][]{1998A&ARv...9....1C,2018Natur.563..657G}. Moreover, the presence of the strong radio jet, with synchrotron emission that extends into the optical regime, has opened a window into the relatively rare class of radio-loud quasars \cite[e.g.,][]{1995ApJ...452L..91B,2005A&A...431..477J,2006ApJ...648..910U}. 

\begin{figure}[t]
\centering
\includegraphics[width=\textwidth]{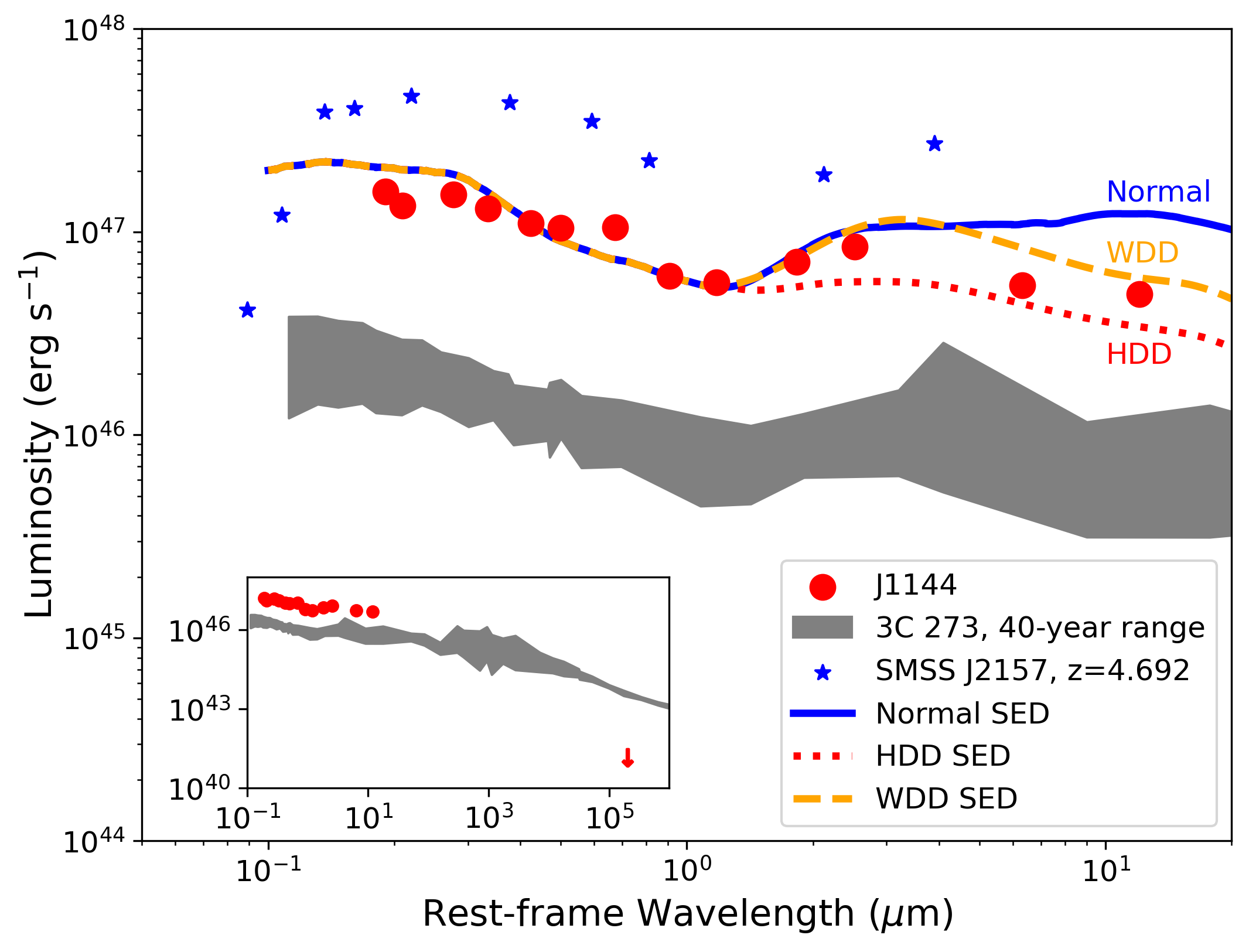}
\caption{Rest-frame SED for J1144 (red circles) compared to the 40-year range of luminosities of 3C~273 (grey shaded region) from \citet{2008A&A...486..411S}, and to SMSS~J2157 (blue stars), the most luminous known quasar. All three sources have been corrected for Galactic extinction. Single-epoch uncertainties for the J1144 photometry are smaller than the symbols. Three quasar templates from \citet{2017ApJ...835..257L} are also shown: normal (solid), hot-dust-deficient (HDD; dotted), and warm-dust-deficient (WDD; dashed). The inset shows an expanded range in order to include the potential J1144 radio association from RACS DR1 as an upper limit (arrow) and to indicate the difference in long-wavelength slope from radio-loud quasars like 3C~273.}
\label{fig:sed}
\end{figure}

In Figure~\ref{fig:sed}, we compare the rest-frame UV-to-IR spectral energy distribution (SED) of J1144 from recent data to the minimum and maximum luminosities of 3C~273 (including synchrotron flares), as observed over a 40-year span\footnote{Data retrieved from the INTEGRAL Science Data Centre (ISDC): \url{http://isdc.unige.ch/3c273/}.} \citep{1999A&AS..134...89T,2008A&A...486..411S}, as well as to SMSS~J2157-3602, the most luminous known quasar in the Universe, at a redshift of $z=4.692$ \citep{2018PASA...35...24W,2020MNRAS.496.2309O}. The J1144 SED is derived from a subset of the photometry presented in Table~\ref{tab:summary}, namely, that of SMSS DR3 ($u$, $v$, $g$, $r$, $i$, $z$), 2MASS ($J, H, K$)\footnote{The $J$ and $K_{\rm s}$ photometry available from DR6 of the VISTA Hemisphere Survey \citep{2013Msngr.154...35M} is little different from the 2MASS data of $\sim20$ years earlier.}, and AllWISE ($W1$, $W2$, $W3$, $W4$). The photometry in Figure~\ref{fig:sed} was corrected for Galactic extinction up to an observed-frame wavelength of 3~$\mu$m using the \citet{2019ApJ...886..108F} extinction curve, a standard $R_{V}$=3.1 Milky Way dust model, and the \citet{1998ApJ...500..525S} E(B-V) values of 0.123, 0.021, and 0.015~mag for J1144, 3C~273, and SMSS~J2157, respectively, with the $\times0.86$ correction factor of \citet{2011ApJ...737..103S}. As with the spectroscopic calibration above, the bandpass details were retrieved from the SVO. The SMSS $u$- and $v$-band photometry were also corrected with the stellar-colour regression method of \citet{2021ApJ...907...68H}, which makes them 0.088 and 0.069~mag fainter, respectively. (The corrections for $g$- and $r$-band are less than 0.01~mag and are therefore omitted.)

Across the entire observed UV/optical range, J1144 has an intrinsic luminosity that is roughly 8 times greater than the brightest observations of 3C~273, and only about 3 times less than the most luminous quasar known. Even with the occasional synchrotron flares elevating the peak luminosities of 3C~273 in the IR, the blazar has remained $\sim5\times$ less luminous than J1144. The inset in Figure~\ref{fig:sed} includes the maximum potential radio luminosity observed for J1144, on the assumption of the RACS DR1 detection being associated with the quasar, illustrating the dramatic difference in the long-wavelength SED of radio-quiet quasars compared to radio-loud sources such as 3C~273. 

In Figure~\ref{fig:sed}, we also show three SED templates\footnote{Retrieved from \url{https://github.com/karlan/AGN_templates}.} \citep{2017ApJ...835..257L} exhibiting different mid-IR dust properties. The "Normal SED" represents the typical broad-line quasar SED, while the hot-dust-deficient (HDD) and warm-dust-deficient (WDD) templates show reduced emission at shorter and longer mid-IR wavelengths, respectively. With the templates anchored to match the J1144 SED near 1~$\mu$m, the {\it WISE} photometry suggests that J1144 is an intermediate case and may be somewhat lacking in dust close to the quasar.

In addition, we note the similarity between the optical spectra of J1144 (Fig.~\ref{fig:spectrum} and \ref{fig:fits}) and 3C~273 \citep{1999A&A...351...31D}, with prominent Balmer lines and comparatively weak [\ion{O}{iii}] emission, suggesting a commonality in their central engines despite the differences in their radio properties. Compared to the sample of bright quasars analysed by BG92, J1144 is on the strong \ion{Fe}{ii}-weak [\ion{O}{iii}] end of their "Eigenvector 1" correlation, although the \ion{Fe}{ii} equivalent width\footnote{We adopt the BG92 method of measuring the \ion{Fe}{ii} flux between 4434 and 4684~\AA.} (EW) of \protect{$\sim40$~\AA}\ is typical of radio-quiet quasars. The ratio of \ion{Fe}{ii}-to-H$\beta$ EWs of 1.5 is at the high end of their distribution, but J1144 does not exhibit the enhanced blue H$\beta$ asymmetry often seen for such sources.

\begin{figure}[t]
\centering
\includegraphics[width=\textwidth]{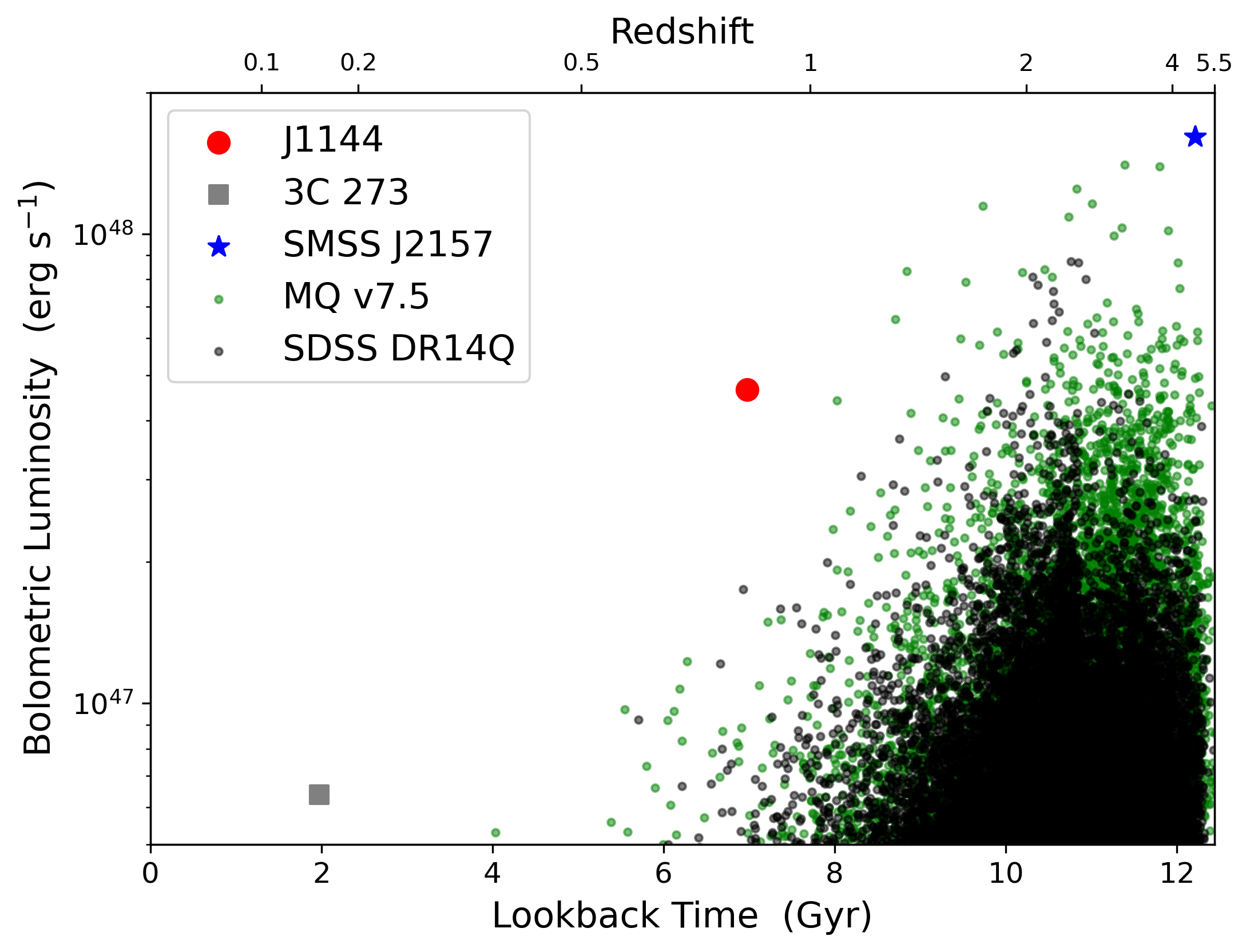}
\caption{Bolometric luminosity of J1144 (large red point), 3C~273 (grey square), and SMSS~J2157 (blue star), compared to sources from the SDSS DR14 quasar catalogue \cite[DR14Q; black points;][]{2020ApJS..249...17R} and Milliquas (MQ; green points), shown as a function of lookback time (bottom axis) and redshift (top axis). No known quasars are as luminous as J1144 in the last 9~Gyr, and J1144 is only a factor of 2 dimmer than the most luminous known quasar, SMSS~J2157.}
\label{fig:lbol_t}
\end{figure}

In Figure~\ref{fig:lbol_t}, we compare the J1144 bolometric luminosity estimated in Section~\ref{sec:lum} to 3C~273, SMSS~J2157, and a large number of quasars, drawn from either the SDSS DR14 quasar catalogue \cite[DR14Q;][]{2020ApJS..249...17R} or Milliquas, as a function of lookback time. 
In order to use a consistent bolometric correction, we estimate the continuum luminosities at 3000~\AA\ for the literature sources. For 3C~273, we use the mean $U$-band flux from the 40-year dataset of the ISDC. DR14Q values use the 3000~\AA\ luminosity tabulated in the catalogue, or, at higher or lower redshifts, respectively, estimate $\log_{10}$ ( $\lambda\,L_{\lambda}(3000)$ )  from the 1350~\AA\ luminosity as $4.887~+~0.89~\log_{10}$ ( $\lambda\,L_{\lambda}(1350)$ ) or from the 5100~\AA\ luminosity as $9.213~+~0.792~\log_{10}$ ( $\lambda\,L_{\lambda}(5100)$ ), based on the best-fit relations from the DR14Q sources having both luminosities estimated. Sources with non-zero 
{\sc QUALITY\_L3000} 
values were excluded. The varied literature sources compiled in Milliquas were cross-matched to the {\it Gaia} catalogue and an empirical scaling from the $R_{p}$ photometry to the 3000~\AA\ luminosity as a function of redshift was applied\footnote{From the {\it Gaia} photometry of DR14Q sources, we estimated a conversion of $\log_{10}$ ($\lambda\,L_{\lambda}(3000) ) = -0.4~R_{p} + (52.9 + 3~\log_{10}(z) - 0.2 z)$ for redshift, $z$. The scaled median absolute deviation (SMAD) of this relation is 0.133~dex, with a median offset of 0.002~dex.}. The Milliquas sample has been restricted to sources with spectroscopic redshifts; cleaned of lensed sources, blazars, and a few spurious objects (including those with $>3\sigma$ parallax or proper motion estimates); and has omitted quasars from SDSS (to avoid duplication). Beyond $z=5.5$, very few Milliquas sources have {\it Gaia} photometry.

As can be seen from Figure~\ref{fig:lbol_t}, J1144 is the most luminous known quasar out to 
$z=1.29$, a lookback time of 8.7~Gyr, beyond which the quasar, HS~2154+2228 \citep{1999A&AS..134..483H},
is the first of a small sample of quasars found to be more luminous, up to the pinnacle of SMSS~J2157. With an extinction-corrected $i$-band magnitude of 14.059~mag (AB), J1144 is nearly 1 full magnitude brighter than the SDSS cutoff at $i_{\rm SDSS}=15$~mag (AB) and nearly 1.5~mag brighter in $M_{i}(z=2)$ than any source actually found in the SDSS DR3 quasar luminosity function at redshifts below $z=0.9$ \citep{2006AJ....131.2766R}.

\section{Discussion}
\label{sec:discussion}

The location of J1144 falls within a small gap in the {\it GALEX} All-Sky Imaging Survey \cite[AIS;][]{2005ApJ...619L...1M}, which explains why it did not appear in DR1 of the UVQS \citep{2016AJ....152...25M}. Utilising $\sim3,000$ known quasars from Milliquas in the same redshift range as J1144, for which both {\it GALEX} and {\it Gaia} photometry exist, we use the $FUV-G$ and $NUV-G$ colours to predict J1144 to have ($FUV$, $NUV$) = (17, 16)~mag (AB), with uncertainties of roughly 1~mag in each band. With $FUV=17$~mag (AB), J1144 would have been amongst the top 10\% of the brightest discoveries by UVQS DR1, but at a redshift 0.2 higher than the rest.

The high luminosity of J1144 also implies a large size for its BLR. Extrapolating the radius--luminosity relation of \citet{2013ApJ...767..149B} to the 5100~\AA\ luminosity of J1144 suggests an H$\beta$-emitting size of $\sim1200$~light-days. With the additional time-dilation factor of $\sim 2$, a reverberation mapping campaign would be a long-term endeavour. However, the angular size of the BLR is expected to be in excess of 100~microarcseconds. Since J1144 has a $K$-band magnitude of 11.9~mag (Vega), its BLR will be well within the reach of the upgraded GRAVITY+ instrument\footnote{See \url{https://www.mpe.mpg.de/ir/gravityplus}.} at ESO's Very Large Telescopes. Thus, it may be possible to measure the Pa$\beta$ dynamics in J1144, comparable to the Pa$\alpha$ measurement for 3C~273 \citep{2018Natur.563..657G}.

Additional studies may make productive use of an exceptionally bright quasar like J1144 as a background source. For example, UV spectroscopy of J1144 may probe the Milky Way's circumgalactic medium \citep{2017ARA&A..55..389T, 2019ApJ...871...35Z, 2021ApJ...912....8B}. 

Previous searches for quasars and other blue objects in the Southern hemisphere have usually not reached as close to the Galactic Plane as J1144, which lies at $b=+18.1^{\circ}$. For example, the Edinburgh-Cape Blue Object Survey \citep{1997MNRAS.287..848S,2016MNRAS.459.4343K} was restricted to $|b|>30^{\circ}$; the Hamburg/ESO quasar survey \citep{1996A&AS..115..227W,2000A&A...358...77W} searched at $|b|>25^{\circ}$; and the Cal\'{a}n-Tololo Survey \citep{1988ASPC....1..410M,1996RMxAA..32...35M} observed to $|b|>20^{\circ}$. Dedicated quasar searches closer to the Galactic Plane \cite[e.g.,][]{2007ApJ...664...64I,2021ApJS..254....6F} may produce samples of objects useful both in their own right and for studies of the gas and dust near the Galactic disk.

Moreover, the discovery power inherent in the recent generation of all-sky surveys like those of {\it Gaia}, {\it WISE}, and {\it eROSITA} motivate a fresh examination of what other bright quasars may have been missed in previous searches across the celestial sphere. A spectroscopic campaign underway at the ANU 2.3m telescope 
has already identified $\sim80$ new, bright quasars (in addition to J1144), some with Galactic latitudes in excess of 60~deg. Thus, after 60 years, it would appear we are finally approaching a complete census of bright quasars, with only the discovery of Changing Look Quasars \citep[CLQ; e.g.,][]{2015ApJ...800..144L} from forthcoming surveys likely to add to the sample.

\begin{acknowledgement}
We thank the anonymous referee for feedback that helped to improve the manuscript. We thank Mara Salvato and Tom Dwelly for fruitful discussions on the X-ray properties, Lisa Crause for helpful input on SpUpNIC, Elaine Sadler for useful discussions on radio counterparts, and David McConnell and Emil Lenc for stimulating discussions on radio astrometry. ABL thanks the rest of the symbiotic star search team, including K.~Mukai, H.~Breytenbach, D.~Buckley, S.~Potter, P.~Woudt, P.~Groot, B.~Paul, N.~Nu\~{n}ez, A.~Howell, M.~Shara, and D.~Zurek, as well as the staff and observers of the American Association of Variable Star Observers and the Astronomical Ring for Access to Spectroscopy. We acknowledge the traditional owners of the land on which the telescopes of Siding Spring Observatory stand, the Kamilaroi people, and pay our respects to their elders, past and present.

CAO was supported by the Australian Research Council (ARC) through Discovery Project DP190100252. ABL and JLS acknowledge support through National Science Foundation (NSF) grant AST-1616646.

The national facility capability for SkyMapper has been funded through ARC LIEF grant LE130100104 from the Australian Research Council, awarded to the University of Sydney, the Australian National University, Swinburne University of Technology, the University of Queensland, the University of Western Australia, the University of Melbourne, Curtin University of Technology, Monash University and the Australian Astronomical Observatory. SkyMapper is owned and operated by The Australian National University's Research School of Astronomy and Astrophysics. The survey data were processed and provided by the SkyMapper Team at ANU. The SkyMapper node of the All-Sky Virtual Observatory (ASVO) is hosted at the National Computational Infrastructure (NCI). Development and support the SkyMapper node of the ASVO has been funded in part by Astronomy Australia Limited (AAL) and the Australian Government through the Commonwealth's Education Investment Fund (EIF) and National Collaborative Research Infrastructure Strategy (NCRIS), particularly the National eResearch Collaboration Tools and Resources (NeCTAR) and the Australian National Data Service Projects (ANDS).

This paper uses observations made at the South African Astronomical Observatory (SAAO).

Based on observations obtained at the Southern Astrophysical Research (SOAR) telescope, which is a joint project of the Minist\'{e}rio da Ci\^{e}ncia, Tecnologia e Inova\c{c}\~{o}es (MCTI/LNA) do Brasil, the US National Science Foundation’s NOIRLab, the University of North Carolina at Chapel Hill (UNC), and Michigan State University (MSU).

This publication makes use of data products from the Wide-field Infrared Survey Explorer, which is a joint project of the University of California, Los Angeles, and the Jet Propulsion Laboratory/California Institute of Technology, and NEOWISE, which is a project of the Jet Propulsion Laboratory/California Institute of Technology. WISE and NEOWISE are funded by the National Aeronautics and Space Administration.

This paper uses data from the VISTA Hemisphere Survey ESO programme ID: 179.A-2010 (PI. McMahon). The VISTA Data Flow System pipeline processing and science archive are described in \citet{2004SPIE.5493..411I}, \citet{2008MNRAS.384..637H} and \citet{2012A&A...548A.119C}.

IRAF was distributed by the National Optical Astronomy Observatory, which was managed by the Association of Universities for Research in Astronomy (AURA) under a cooperative agreement with the NSF.

We acknowledge use of the International Centre for Radio Astronomy Research (ICRAR) Cosmology Calculator written by Aaron Robotham and Joseph Dunne, and available at \url{https://cosmocalc.icrar.org}.

This research has made use of the NASA/IPAC Extragalactic Database (NED), which is funded by the National Aeronautics and Space Administration and operated by the California Institute of Technology.
This research has made use of the SIMBAD database,
operated at CDS, Strasbourg, France.
This research has made use of "Aladin sky atlas" developed at CDS, Strasbourg Observatory, France .

SuperCOSMOS Sky Survey material is based on photographic data originating from the UK, Palomar and ESO Schmidt telescopes and is provided by the Wide-Field Astronomy Unit, Institute for Astronomy, University of Edinburgh.

This work has made use of data from the Asteroid Terrestrial-impact Last Alert System (ATLAS) project. The Asteroid Terrestrial-impact Last Alert System (ATLAS) project is primarily funded to search for near earth asteroids through NASA grants NN12AR55G, 80NSSC18K0284, and 80NSSC18K1575; byproducts of the NEO search include images and catalogs from the survey area. This work was partially funded by Kepler/K2 grant J1944/80NSSC19K0112 and HST GO-15889, and STFC grants ST/T000198/1 and ST/S006109/1. The ATLAS science products have been made possible through the contributions of the University of Hawaii Institute for Astronomy, the Queen’s University Belfast, the Space Telescope Science Institute, the South African Astronomical Observatory, and The Millennium Institute of Astrophysics (MAS), Chile.

Funding for the {\it TESS} mission is provided by NASA’s Science Mission directorate. This paper includes data collected by the {\it TESS} mission, which are publicly available from the Mikulski Archive for Space Telescopes (MAST). This research made use of Lightkurve, a Python package for Kepler and TESS data analysis \citep{2018ascl.soft12013L}. This research made use of Astropy,\footnote{See \url{http://www.astropy.org}.} a community-developed core Python package for Astronomy \citep{2013A&A...558A..33A, 2018AJ....156..123A}. This research made use of the astroquery \citep{2019AJ....157...98G} and Astrocut \citep{2019ascl.soft05007B} packages for Python.

This project used data obtained with the Dark Energy Camera (DECam), which was constructed by the Dark Energy Survey (DES) collaboration. Funding for the DES Projects has been provided by the US Department of Energy, the US National Science Foundation, the Ministry of Science and Education of Spain, the Science and Technology Facilities Council of the United Kingdom, the Higher Education Funding Council for England, the National Center for Supercomputing Applications at the University of Illinois at Urbana-Champaign, the Kavli Institute for Cosmological Physics at the University of Chicago, Center for Cosmology and Astro-Particle Physics at the Ohio State University, the Mitchell Institute for Fundamental Physics and Astronomy at Texas A\&M University, Financiadora de Estudos e Projetos, Funda\c{c}\~{a}o Carlos Chagas Filho de Amparo \`{a} Pesquisa do Estado do Rio de Janeiro, Conselho Nacional de Desenvolvimento Cient\'{i}fico e Tecnol\'{o}gico and the Minist\'{e}rio da Ci\^{e}ncia, Tecnologia e Inova\c{c}\~{a}o, the Deutsche Forschungsgemeinschaft and the Collaborating Institutions in the Dark Energy Survey.
The Collaborating Institutions are Argonne National Laboratory, the University of California at Santa Cruz, the University of Cambridge, Centro de Investigaciones En\'{e}rgeticas, Medioambientales y Tecnol\'{o}gicas–Madrid, the University of Chicago, University College London, the DES-Brazil Consortium, the University of Edinburgh, the Eidgen\"{o}ssische Technische Hochschule (ETH) Z\"{u}rich, Fermi National Accelerator Laboratory, the University of Illinois at Urbana-Champaign, the Institut de Ci\`{e}ncies de l’Espai (IEEC/CSIC), the Institut de F\/{i}sica d’Altes Energies, Lawrence Berkeley National Laboratory, the Ludwig-Maximilians Universit\"{a}t M\"{u}nchen and the associated Excellence Cluster Universe, the University of Michigan, the National Science Foundation's NOIRLab, the University of Nottingham, the Ohio State University, the OzDES Membership Consortium, the University of Pennsylvania, the University of Portsmouth, SLAC National Accelerator Laboratory, Stanford University, the University of Sussex, and Texas A\&M University.
Based on observations at Cerro Tololo Inter-American Observatory, a program of NOIRLab (NOIRLab Prop. ID 2017A-0260; PI: M.\ Soares-Santos; and Prop. ID 2019A-0272; PI: A.\ Zenteno), which is managed by AURA under a cooperative agreement with the NSF.
This research draws upon DECam data as distributed by the Astro Data Archive at NSF's NOIRLab. NOIRLab is managed by AURA under a cooperative agreement with the NSF.

This work has made use of data from the European Space Agency (ESA) mission
{\it Gaia} (\url{https://www.cosmos.esa.int/gaia}), processed by the {\it Gaia}
Data Processing and Analysis Consortium (DPAC,
\url{https://www.cosmos.esa.int/web/gaia/dpac/consortium}). Funding for the DPAC
has been provided by national institutions, in particular the institutions
participating in the {\it Gaia} Multilateral Agreement.

The DASCH project at Harvard is grateful for partial support from NSF grants AST-0407380, AST-0909073, and AST-1313370.

\end{acknowledgement}

\bibliography{bib}



\end{document}